\documentclass[journal=mamobx,manuscript=article]{achemso}

\usepackage{amsmath,amssymb,amsfonts,bm}
\usepackage{graphicx}
\usepackage{booktabs}
\usepackage{siunitx}
\usepackage{amsmath,amssymb,amsfonts,bm}

\newcommand{\NLI}{\mathrm{NLI}}

\title{\bfseries Delayed Arm Retraction Controls the Nonlinear Oscillatory Response of Long-Chain-Branched Polymer Melts}

\author{Dario Nichetti}
\affiliation{Rheonic Lab, Via Quadelle 2C, 26012 Castelleone (CR), Italy}
\email{dario.nichetti@rheoniclab.com}

\author{Alessio Zaccone}
\affiliation{Department of Physics ``Aldo Pontremoli'',
University of Milan,
Via Celoria 16,
20133 Milan,
Italy}
\email{alessio.zaccone@unimi.it}

\begin{document}
\maketitle

\begin{abstract}
Long-chain branching profoundly modifies the nonlinear oscillatory response of entangled polymer melts by introducing arm-retraction pathways absent in linear polymers. We present a molecular tube theory that explains the characteristic maximum of the Nonlinearity Index (NLI) observed experimentally in long-chain-branched polymers. The theory extends the recently developed nonlinear tube-orientation description of linear polymers by incorporating branch-point force transmission and delayed arm retraction. The backbone initially develops nonlinear orientation as in the corresponding linear polymer, whereas long-arm retraction subsequently relaxes the stored branch-point tension and progressively erases backbone orientational memory. This competition produces a characteristic NLI maximum followed by a post-peak decay. The theory predicts two distinct nonlinear regimes corresponding to sparse and dense long-chain branching and introduces an architecture parameter governing the height and width of the nonlinear peak. The resulting framework provides a molecular interpretation of nonlinear Fourier rheology and directly links the nonlinear harmonic response to polymer architecture.
\end{abstract}


\section{Introduction}
Long-chain branching is one of the most sensitive molecular variables controlling the nonlinear rheology of entangled polymers \cite{Read_Science,ReadMcLeish}. A small number of long arms can strongly change strain hardening, stress overshoots, extensional viscosity, relaxation spectra, and nonlinear oscillatory shear response because branch points couple the motion of different tube segments and introduce arm-retraction pathways that are absent in linear chains. Activated arm retraction along the confining
tube according to the Milner--McLeish theory of branched polymer dynamics \cite{Milner1998,Milner2001} extends the original Doi--Edwards tube framework to long-chain-branched architectures \cite{DoiEdwards1986,RubinsteinColby2003,McLeish2002,Hoyle2014}. The resulting nonlinear response is not simply a shifted version of the linear-polymer response; it reflects competition between backbone orientation, branch-point motion, arm retraction, chain stretch, and convective constraint release (CCR) \cite{Marrucci1996,MarrucciIanniruberto1996,LikhtmanMcLeish2002,LikhtmanGraham2003}.

Fourier-transform rheology and large-amplitude oscillatory shear have become standard tools for resolving nonlinear responses in polymer melts \cite{HassagerVlassopoulos2008}, blends and elastomers because the stress waveform contains higher harmonics that encode intra-cycle structural relaxation processes \cite{Wilhelm2002,Hyun2011,Ewoldt2008,Macaubas2005}. The Nonlinearity Index (NLI), defined from the weighted higher odd elastic harmonics normalized by the first-harmonic storage modulus, provides a compact scalar measure of the nonlinear elastic distortion \cite{NichettiScacchi2025,ScacchiNichetti2025AJOP}. In linear and weakly branched entangled polymers, the NLI often grows approximately monotonically with strain amplitude over the accessible experimental window. By contrast, long-chain-branched polymers may show a qualitatively different response: the NLI increases after the onset of nonlinearity, reaches a maximum, and then decreases at larger strain amplitude \cite{NichettiScacchi2025,ScacchiNichetti2025AJOP}.

The central question addressed in this paper is therefore not merely how to fit the NLI curve, but why long-chain branching produces a maximum in the first place. We propose that the maximum arises from a molecular competition. At the beginning of nonlinear deformation, the backbone tube segment is oriented by the imposed oscillatory flow and the stress waveform becomes increasingly distorted. Branch points act initially as temporary anchors, allowing nonlinear backbone orientation to build up. At larger strain amplitudes, however, the long arms begin to retract and relax the branch-point force balance during the cycle. This arm relaxation, assisted by CCR and tube dilation, destroys part of the stored backbone orientation and reduces the nonlinear elastic harmonic content. The NLI maximum marks the crossover between these two regimes. This interpretation is consistent with the broader picture based on nonaffine elasticity \cite{Zaccone2011,Zaccone2023book} developed in Refs.~\cite{NichettiZaccone2026Nonaffine,NichettiZaccone2026Constraint}. In that framework, the NLI is interpreted as a Fourier-resolved measure of dynamic nonaffinity: the first harmonic measures the residual phase-locked elastic response, whereas the higher harmonics quantify strain-dependent nonaffine relaxation channels. The present work extends that viewpoint to long-chain-branched polymers by identifying arm retraction and branch-point relaxation as architecture-specific nonaffine relaxation mechanisms. Thus, the decrease of the NLI after its maximum is not simply a loss of signal, but the signature of a progressive loss of coherent backbone orientation caused by branch-mediated relaxation.
 
This mechanism is distinct from the saturation mechanism in linear entangled polymers. In the constraint-limited tube-orientation picture, the NLI is bounded because the shear component of the tube-orientation tensor cannot exceed its geometrical limit. In long-chain-branched polymers, the nonlinear response can be suppressed before this orientational ceiling is reached because branch-point relaxation and arm retraction compete with backbone orientation. Thus the NLI maximum in branched polymers is an architectural effect, not simply an experimental manifestation of the universal orientational bound.

The purpose of the present work is to formulate this picture quantitatively. We develop a minimal molecular model based on the variables $Z_{bb}$, the number of entanglements between neighboring branch points along the backbone, and $Z_a$, the number of entanglements in a long arm. A branch-point force balance gives an architecture factor $\phi_b=Z_{bb}/(Z_{bb}+f_bZ_a)$ that reduces the attainable backbone orientation. This factor leads directly to a prediction for the peak NLI and provides a molecular criterion for distinguishing sparse and dense long-arm branching. The linear polymer is recovered exactly by setting $Z_a=0$.

\section{Nonlinear harmonic measure}

We consider oscillatory shear
\begin{equation}
\gamma(t)=\gamma_0\sin(\omega t),
\qquad
\dot\gamma(t)=\gamma_0\omega\cos(\omega t),
\end{equation}
where $\gamma_0$ is the strain amplitude and $\omega$ is the angular frequency. The elastic stress is expanded in odd sine harmonics,
\begin{equation}
\tau_E(t)=\sum_{n=0}^{\infty}\tau'_{2n+1}\sin[(2n+1)\omega t].
\end{equation}
The first-harmonic storage modulus is
\begin{equation}
G'_1(\gamma_0)=\frac{\tau'_1}{\gamma_0},
\end{equation}
and the nonlinear elastic modulus is defined as
\begin{equation}
G'_{NL}(\gamma_0)=\sum_{n=1}^{\infty}(2n+1)\frac{\tau'_{2n+1}}{\gamma_0}.
\end{equation}
The Nonlinearity Index is then
\begin{equation}
\label{eq:NLI_def}
\NLI(\gamma_0)=-\frac{G'_{NL}(\gamma_0)}{G'_1(\gamma_0)}.
\end{equation}
The minus sign is used because the nonlinear elastic contribution is negative for strain-softening nonlinear elasticity, whereas the reported NLI is positive.

Within a molecular interpretation, $G'_1$ represents the coherent first-harmonic storage associated with phase-locked tube orientation, whereas $G'_{NL}$ measures the cumulative higher-harmonic distortion produced by strain-dependent relaxation of the tube memory. The NLI therefore measures how much elastic response has been transferred from coherent first-harmonic storage into nonlinear harmonic modes.

\section{Representative long-chain-branched molecule}

We describe the branched polymer by a representative backbone segment and one or more long arms attached to a branch point. The total number of entanglements of a corresponding linear chain is
\begin{equation}
Z=\frac{M}{M_e},
\end{equation}
where $M$ is molecular weight and $M_e$ is the entanglement molecular weight. The backbone segment between consecutive branch points is characterized by
\begin{equation}
Z_{bb}=\frac{M_{bb}}{M_e},
\end{equation}
and a long arm is characterized by
\begin{equation}
Z_a=\frac{M_a}{M_e}.
\end{equation}
The present theory focuses on long arms,
\begin{equation}
Z_a\gtrsim 3,
\end{equation}
so that the arms are themselves entangled and cannot relax as simple Rouse side groups. Their dominant relaxation mechanism is arm retraction, whose characteristic time increases rapidly with arm length.

The architecture is controlled primarily by the ratio $Z_a/Z_{bb}$. Large $Z_a/Z_{bb}$ corresponds to long arms attached to relatively short backbone spans and therefore to dense or strongly constraining long-chain branching. Small $Z_a/Z_{bb}$ corresponds to sparse long-chain branching, where the backbone span between branch points is sufficiently long to preserve a larger fraction of the linear-chain orientation.

\section{Backbone stress and branch-point force balance}

The elastic tube stress is assigned to the backbone tube segment because the backbone carries the long-time orientational memory. We write
\begin{equation}
\label{eq:stress_bb}
\tau_E(t)=G_e \lambda^2(t) S_{xy}^{(bb)}(t),
\end{equation}
where $G_e$ is the entanglement modulus, $\lambda$ is the chain stretch ratio, and $S_{xy}^{(bb)}$ is the shear component of the backbone orientation tensor. At large strain amplitude, stretch may approach a saturated state, and further nonlinear evolution is dominated primarily by orientation loss and constraint renewal. In this orientation-controlled regime,
\begin{equation}
\tau_E(t)\simeq G_e S_{xy}^{(bb)}(t).
\end{equation}

The T-branch is not merely a geometrical junction connecting the backbone and the long arms. It acts as a molecular force-transfer node through which the orientational stress generated in the backbone is transmitted to the arms. During oscillatory shear, the backbone tends to orient in the flow direction, while the long arms resist this displacement because they remain temporarily trapped within their respective confining tubes. Consequently, the imposed macroscopic deformation is converted into a local branch-point deformation and an associated entropic restoring force. This local force balance determines the fraction of backbone orientation that survives in the branched molecule. A minimal entropic force balance gives the fraction of orientation stored on the backbone as
\begin{equation}
\label{eq:phib_single}
\phi_b=\frac{Z_{bb}}{Z_{bb}+Z_a}.
\end{equation}
For an effective branch functionality $f_b$, the resisting arm contribution is multiplied by $f_b$, giving
\begin{equation}
\label{eq:phib_general}
\phi_b=\frac{Z_{bb}}{Z_{bb}+f_b Z_a}.
\end{equation}
The architecture factor therefore does not simply reduce the backbone orientation. It represents the efficiency with which orientational stress generated in the backbone is transmitted through the T-branch before being relaxed by the long arms.
Equation \eqref{eq:phib_general} is the simplest normalized expression satisfying the limiting requirements of the molecular force balance. It recovers the linear-chain limit $\phi_b\rightarrow1$ for $Z_a\rightarrow0$ and predicts complete suppression of backbone orientation when the restoring force exerted by the arms dominates the backbone elasticity ($f_bZ_a\gg Z_{bb}$).
The backbone orientation is therefore reduced relative to the corresponding linear-chain orientation:
\begin{equation}
\label{eq:Sbb_reduced}
S_{xy}^{(bb)}\simeq \phi_b S_{xy}^{lin}.
\end{equation}

The physical meaning is direct. If $Z_{bb}\gg f_bZ_a$, the branch point weakly perturbs backbone orientation. If $f_bZ_a\gg Z_{bb}$, the arms dominate the force balance and the backbone orientation is strongly suppressed. This is the molecular origin of the lower NLI peak in dense long-arm-branched polymers.

Because the primitive-path tangent is a unit vector, the backbone shear orientation remains bounded. The linear-chain limit gives $|S_{xy}^{lin}|\leq 1/2$, while the branched backbone satisfies
\begin{equation}
\label{eq:Sbb_bound}
|S_{xy}^{(bb)}|\leq \frac{1}{2}\phi_b.
\end{equation}
This bound is central to the prediction of the maximum NLI in branched architectures.

\section{CCR, stretch, and arm-retraction memory}

The backbone orientation evolves through affine deformation, reptative relaxation, and deformation-induced loss of tube memory by CCR. A compact scalar representation is
\begin{equation}
\frac{dS_{xy}^{(bb)}}{dt}=\dot\gamma(t)\mathcal A(t)-\frac{S_{xy}^{(bb)}}{\tau_d^{\textrm{eff}}},
\end{equation}
where $\mathcal A(t)$ is an order-one orientational source term and $\tau_d^{\textrm{eff}}$ is the effective disengagement time. The effective relaxation rate is written as
\begin{equation}
\label{eq:tau_eff}
\frac{1}{\tau_d^{\textrm{eff}}}=\frac{1}{\tau_d}+\frac{1}{\tau_{CCR}},
\end{equation}
with
\begin{equation}
\label{eq:ccr_rate}
\frac{1}{\tau_{CCR}(t)}=\beta_{CCR}|\dot\gamma(t)|=\beta_{CCR}\gamma_0\omega|\cos(\omega t)|.
\end{equation}
Thus CCR becomes more important as the strain amplitude increases and is strongest near the zero-strain crossings, where the shear rate is largest.

The stretch variable satisfies a minimal equation of the form
\begin{equation}
\frac{d\lambda}{dt}=\lambda S_{xy}^{(bb)}(t)\dot\gamma(t)-\frac{\lambda-1}{\tau_R^{\textrm{eff}}}-\mathcal F_{sat}(\lambda),
\end{equation}
where $\tau_R^{\textrm{eff}}$ is an effective stretch relaxation time and $\mathcal F_{sat}$ prevents unlimited chain extension. A useful saturation term is
\begin{equation}
\mathcal F_{sat}(\lambda)=\frac{1}{\tau_R^{\textrm{eff}}}\frac{\lambda^2-1}{\lambda_{max}^2-\lambda^2}\lambda.
\end{equation}
CCR also accelerates stretch relaxation,
\begin{equation}
\frac{1}{\tau_R^{\textrm{eff}}}=\frac{1}{\tau_R}+\beta_s |\dot\gamma(t)|.
\end{equation}
At high strain amplitude the stretch relaxation becomes fast during part of the cycle, so the strong nonlinear regime can be treated, to leading order, as governed by orientation loss and arm-retraction memory.

The long arms relax predominantly by thermally activated arm retraction within the confining tube \cite{Milner1998,Milner2001}. For deeply entangled arms, Milner--McLeish tube theory predicts that the probability of an arm retaining its original tube orientation follows an asymptotic stretched-exponential decay rather than a single exponential. Accordingly, we adopt the survival probability
\begin{equation}
\label{eq:arm_survival}
\Psi_a(t)=
\exp\!\left[
-\left(
\frac{t}{\tau_a}
\right)^{1/2}
\right],
\end{equation}
where the characteristic arm-retraction time is
\begin{equation}
\label{eq:arm_time}
\tau_a\simeq \tau_e Z_a^{3/2}\exp\left(\frac{3}{2}Z_a\right).
\end{equation}
The exponent $1/2$ in Eq. \eqref{eq:arm_survival} is is the asymptotic prediction of the Milner--McLeish \cite{Milner1998,Milner2001}
arm-retraction theory for highly entangled branches and is therefore not treated
as an adjustable parameter. Physically, the stretched-exponential form reflects the broad spectrum of thermally activated arm-retraction trajectories available within the confining tube.
The backbone memory kernel is then approximated as the product of CCR-modified reptation memory and arm survival,
\begin{equation}
\label{eq:memory_kernel}
M_{br}(t)=\exp\left(-\frac{t}{\tau_d^{\textrm{eff}}}\right)\exp\left[-\left(\frac{t}{\tau_a}\right)^{1/2}\right].
\end{equation}
The backbone orientation in oscillatory shear follows from the convolution
\begin{equation}
\label{eq:convolution}
S_{xy}^{(bb)}(t)=\phi_b\int_0^{\infty}M_{br}(t')\dot\gamma(t-t')\,dt'.
\end{equation}

The backbone memory kernel in Eq.~\eqref{eq:convolution} combines the usual reptation and CCR relaxation
with the additional loss of orientational memory caused by delayed arm retraction. During
the early stages of deformation, the long arms remain effectively trapped within their
confining tubes and the backbone develops nonlinear orientation. As arm retraction becomes
active, the branch-point restoring force progressively relaxes, shortening the lifetime of
backbone orientational memory and reducing the nonlinear harmonic content. This molecular
picture is developed quantitatively in Section~"Maxmimum NLI and delayed arm-retraction crossover" below.

Approximating the convolution in Eq.\eqref{eq:convolution} by its dominant Fourier component immediately gives the relation:
\begin{equation}
\label{eq:Sbb_freq}
S_{xy}^{(bb)}\propto \phi_b\frac{\gamma_0\omega}{\omega+1/\tau_d^{\textrm{eff}}+(\omega/\tau_a)^{1/2}}.
\end{equation}
At low frequency, arms retract within a cycle and suppress the nonlinear harmonic content. At high frequency, arms cannot fully retract and branch points behave more like persistent anchors.

\section{Onset strain and architecture shift}

The onset of nonlinear response occurs when CCR and tube dilation first modify the orientation memory during one cycle. For a linear chain, we write
\begin{equation}
\label{eq:gamma_c_lin}
\gamma_c^{lin}=C_c Z^{-1/2}=C_c\left(\frac{M_e}{M}\right)^{1/2},
\end{equation}
where $C_c$ depends on the operational definition of the linear-regime limit.

For a branched molecule, the effective orientable backbone fraction is reduced by $\phi_b$. To reach the same nonlinear orientational distortion, the imposed strain must therefore be larger by the factor $\phi_b^{-1/2}$. Thus
\begin{equation}
\label{eq:gamma_c_br_general}
\gamma_c^{br}=\gamma_c^{lin}\phi_b^{-1/2}.
\end{equation}
Substituting Eq. \eqref{eq:phib_general} gives
\begin{equation}
\label{eq:gamma_c_br}
\gamma_c^{br}=C_c\left(\frac{M_e}{M}\right)^{1/2}\left(1+\frac{f_bZ_a}{Z_{bb}}\right)^{1/2}.
\end{equation}
The linear limit is immediate: setting $Z_a=0$ gives $\phi_b=1$ and $\gamma_c^{br}\rightarrow \gamma_c^{lin}$.

\section{Maximum NLI and delayed arm-retraction crossover}

We use the nonlinear crossover form introduced in Ref. \cite{NichettiZaccone2026Nonaffine} for the backbone response of entangled polymers,
\begin{equation}
\label{eq:nli_paper1}
\NLI^{(1)}(\theta)
=
N_{\max}
\frac{\theta^m}{1+\theta^m},
\qquad
\theta=\frac{\gamma_0}{\gamma_c}.
\end{equation}
This expression is Eq.~(83) in Ref.~\cite{NichettiZaccone2026Nonaffine}. Here \(N_{\max}\) is the nonlinear saturation level reached by the corresponding backbone response, while \(m\) is the effective crossover exponent controlling the growth of higher harmonics.
Equation \eqref{eq:nli_paper1} is not an independent constitutive equation but an analytical representation of the nonlinear backbone response obtained previously. Its purpose is to obtain simple analytical predictions for the architecture dependence introduced below.

Long-chain branching is introduced as an additional relaxation channel acting on this backbone response. Before efficient arm retraction occurs, the long arms behave as transient branch-point constraints and the branched polymer follows the same orientational buildup as the corresponding linear backbone. We therefore write
\begin{equation}
\label{eq:nli_branch_factorized}
\NLI^{br}(\theta)
=
\NLI^{(1)}(\theta)\mathcal H_a(\theta),
\end{equation}
where \(\mathcal H_a\) is a branch-memory survival factor.

The delayed arm-retraction factor is
\begin{equation}
\label{eq:Harm_delayed}
\mathcal H_a(\theta)
=
\begin{cases}
1, & \theta\leq\theta_m,\\[6pt]
\left[
1+
B_a
\left(
\dfrac{\theta-\theta_m}{\theta_a}
\right)^p
\right]^{-q},
& \theta>\theta_m,
\end{cases}
\end{equation}
with
\begin{equation}
\label{eq:Ba_definition}
B_a=\frac{f_bZ_a}{Z_{bb}}.
\end{equation}
The algebraic form adopted here, Eq.~\eqref{eq:Harm_delayed}, is motivated by the asymptotic
arm-retraction survival probability of Milner and McLeish
\cite{Milner1998,Milner2001,McLeish2002}, but is written as a simple
Pad\'e-type crossover function suitable for analytical treatment.
Within the Milner--McLeish tube theory of branched polymers, relaxation of
an entangled arm occurs through activated retraction along its confining
tube, leading to a characteristic survival probability for branch-point
memory \cite{Milner1998,Milner2001,McLeish2002}. In this theory, the survival
probability of an entangled arm is asymptotically
\begin{equation}
\Psi_a(t)
=
\exp\left[-\left(\frac{t}{\tau_a}\right)^{1/2}\right],
\end{equation}
where \(\tau_a\) is the arm-retraction time. Under LAOS, the relevant time
available for branch-point relaxation during a cycle decreases as the
strain amplitude increases beyond the activation threshold \(\theta_m\).
Equivalently, the accumulated branch-point loading grows with the excess
reduced strain \(\theta-\theta_m\). We therefore introduce the dimensionless
activation variable
\begin{equation}
X_a(\theta)
=
B_a
\left(
\frac{\theta-\theta_m}{\theta_a}
\right)^p,
\qquad
\theta>\theta_m,
\end{equation}
where \(B_a=f_bZ_a/Z_{bb}\) measures the strength of the arm-mediated
relaxation channel. The present algebraic form should therefore be regarded as a compact
crossover approximation to the Milner--McLeish arm-retraction survival
function \cite{Milner1998,Milner2001}, chosen to preserve the correct small-deformation limit and the
large-strain tube-dilation asymptote.

A convenient Pad\'e-type representation of the survival of backbone
orientational memory is then
\begin{equation}
\mathcal H_a(\theta)
=
\left[1+X_a(\theta)\right]^{-q}.
\end{equation}
For small excess deformation, \(X_a\ll1\), this gives
\begin{equation}
\mathcal H_a\simeq 1-qX_a,
\end{equation}
so branch memory is only weakly reduced immediately after the onset of arm
retraction. For large deformation, \(X_a\gg1\),
\begin{equation}
\mathcal H_a\sim X_a^{-q}
\sim
\left(\theta-\theta_m\right)^{-pq},
\end{equation}
which gives the required power-law post-peak decay. Choosing
\begin{equation}
pq=\frac45
\end{equation}
recovers the tube-dilation asymptote used in nonlinear tube theories.
Thus Eq.~\eqref{eq:Harm_delayed} should be interpreted as a compact
crossover approximation to the arm-retraction survival memory, constrained
by the correct small-deformation limit and by the large-strain
tube-dilation decay.

Combining Eqs.~\eqref{eq:nli_paper1}--\eqref{eq:Harm_delayed} gives
\begin{equation}
\label{eq:finalNLI}
\NLI^{br}(\theta)
=
\begin{cases}
N_{\max}
\dfrac{\theta^m}{1+\theta^m},
& \theta\leq\theta_m,\\[12pt]
N_{\max}
\dfrac{\theta^m}{1+\theta^m}
\left[
1+
B_a
\left(
\dfrac{\theta-\theta_m}{\theta_a}
\right)^p
\right]^{-q},
& \theta>\theta_m.
\end{cases}
\end{equation}

For sufficiently large strain amplitude, the post-peak branch-memory factor gives
\begin{equation}
\NLI^{br}\sim \theta^{-pq}.
\end{equation}
The tube-dilation asymptote is recovered by choosing
\begin{equation}
pq=\frac{4}{5}.
\end{equation}

The physical interpretation of Eq.~\eqref{eq:finalNLI} is direct. The factor \(\NLI^{(1)}\) describes the nonlinear backbone response already obtained from the nonaffine tube-orientation theory. The factor \(\mathcal H_a\) describes the delayed loss of backbone orientational memory caused by arm retraction. Thus, branching does not introduce a new mechanism for generating nonlinear orientation; it introduces an architecture-dependent pathway for relaxing that orientation once the T-branch force exceeds the arm-retraction threshold.

For sparse long-chain branching, the number of entangled arms attached to each backbone segment remains small and the backbone therefore retains essentially the same nonlinear orientational buildup as the corresponding linear polymer over a broad strain interval. Arm retraction is activated only at larger deformation, allowing the NLI maximum to exceed unity before the branch-mediated relaxation channel becomes dominant. As the density of long-chain branching increases, the delayed relaxation sets in at progressively smaller reduced strain, reducing both the height and the width of the nonlinear peak.

\section{Comparison with experimental NLI curves}

\begin{figure}[htbp]
\centering
\includegraphics[width=0.82\textwidth]{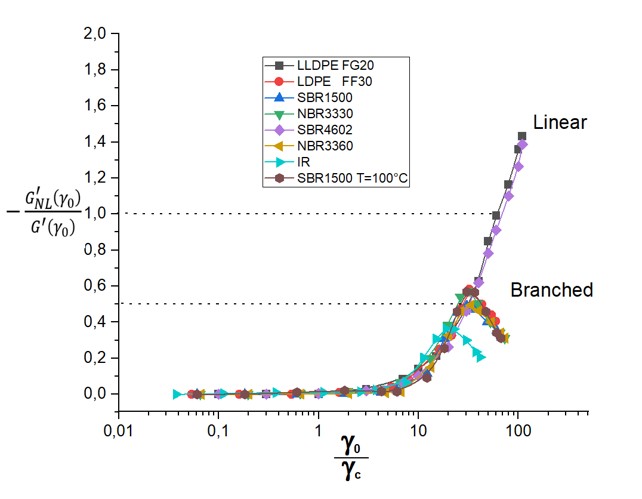}
\caption{
Experimental Nonlinearity Index as a function of reduced strain amplitude $\gamma_0/\gamma_c$ for representative linear and long-chain-branched polymers compiled from Refs.~\cite{NichettiScacchi2025,ScacchiNichetti2025AJOP}. The linear materials (LLDPE FG20, LDPE FP30 and IR) display monotonic growth over the measured strain range. In contrast, the branched elastomers (SBR1500, SBR4602, NBR3330 and NBR3360) exhibit a characteristic maximum, typically around NLI $\approx0.5$--$0.6$, followed by a decrease at larger strain amplitudes. Within the present theory, the initial increase reflects nonlinear buildup of backbone orientation, whereas the post-peak decay arises from arm retraction assisted by convective constraint release and tube dilation.
}
\label{fig:experimental}
\end{figure}

The experimental behavior motivating the model is shown in Fig.~\ref{fig:experimental}, which compiles representative NLI data for both linear and long-chain-branched polymers obtained in previous studies by Nichetti and co-workers \cite{NichettiScacchi2025,ScacchiNichetti2025AJOP}. The linear polymers (LLDPE, LDPE and IR) exhibit a monotonic increase of the NLI over the experimentally accessible reduced strain range. By contrast, the long-chain-branched elastomers (SBR and NBR grades with different branching architectures) develop a pronounced maximum followed by a decrease at larger reduced strain amplitudes. This qualitative difference motivates the molecular mechanism developed in the present work.

Although these materials differ chemically, they exhibit the same qualitative architectural distinction: polymers without long-chain branching display monotonic NLI growth, whereas polymers containing long entangled branches develop a peak followed by a decay. The present theory aims to explain this universal architectural trend rather than chemistry-specific details.


The experimental curves shown in Fig.~1 exhibit a systematic reduction of both the height and the width of the NLI maximum as the degree of long-chain branching increases. Within the present theory, this behavior is controlled by the branching parameter
\begin{equation}
B_a=\frac{f_bZ_a}{Z_{bb}},
\end{equation}
which enters the delayed arm-retraction factor in Eqs.~(7.3)--(7.5). Increasing $B_a$ strengthens the branch-mediated relaxation pathway once arm retraction becomes active, thereby producing progressively lower and broader NLI maxima. Conversely, small values of $B_a$ correspond to weak branch-mediated relaxation and recover the nearly monotonic behavior characteristic of linear polymers.

The model therefore predicts that the measured NLI peak provides direct information on the relative importance of arm entanglements compared with the orientable backbone span between neighbouring branch points. Rather than being interpreted as an empirical fitting parameter, the peak height becomes a molecular indicator of the competition between backbone orientation and delayed arm retraction.

A key observation is that the maximum in the branched-polymer curves occurs at a similar reduced deformation. Define
\begin{equation}
\theta=\frac{\gamma_0}{\gamma_c},
\qquad
\theta_m=\frac{\gamma_m}{\gamma_c}.
\end{equation}
The reduced deformation should not be interpreted merely as a normalized strain. Rather, within the present molecular picture it represents a normalized T-branch loading coordinate. The onset strain $\gamma_c$ normalizes the deformation required to activate nonlinear backbone orientation at the T-branch, while $\gamma_m$ marks the subsequent reduced deformation at which the same branch-point force activates dominant arm retraction. The approximate collapse of the experimental maxima therefore suggests that the arm-retraction-dominated regime is reached after accumulating nearly the same relative T-branch deformation, rather than the same absolute strain.


\section{Predicted architecture dependence}

\begin{figure}[htbp]
\centering
\includegraphics[width=0.82\textwidth]{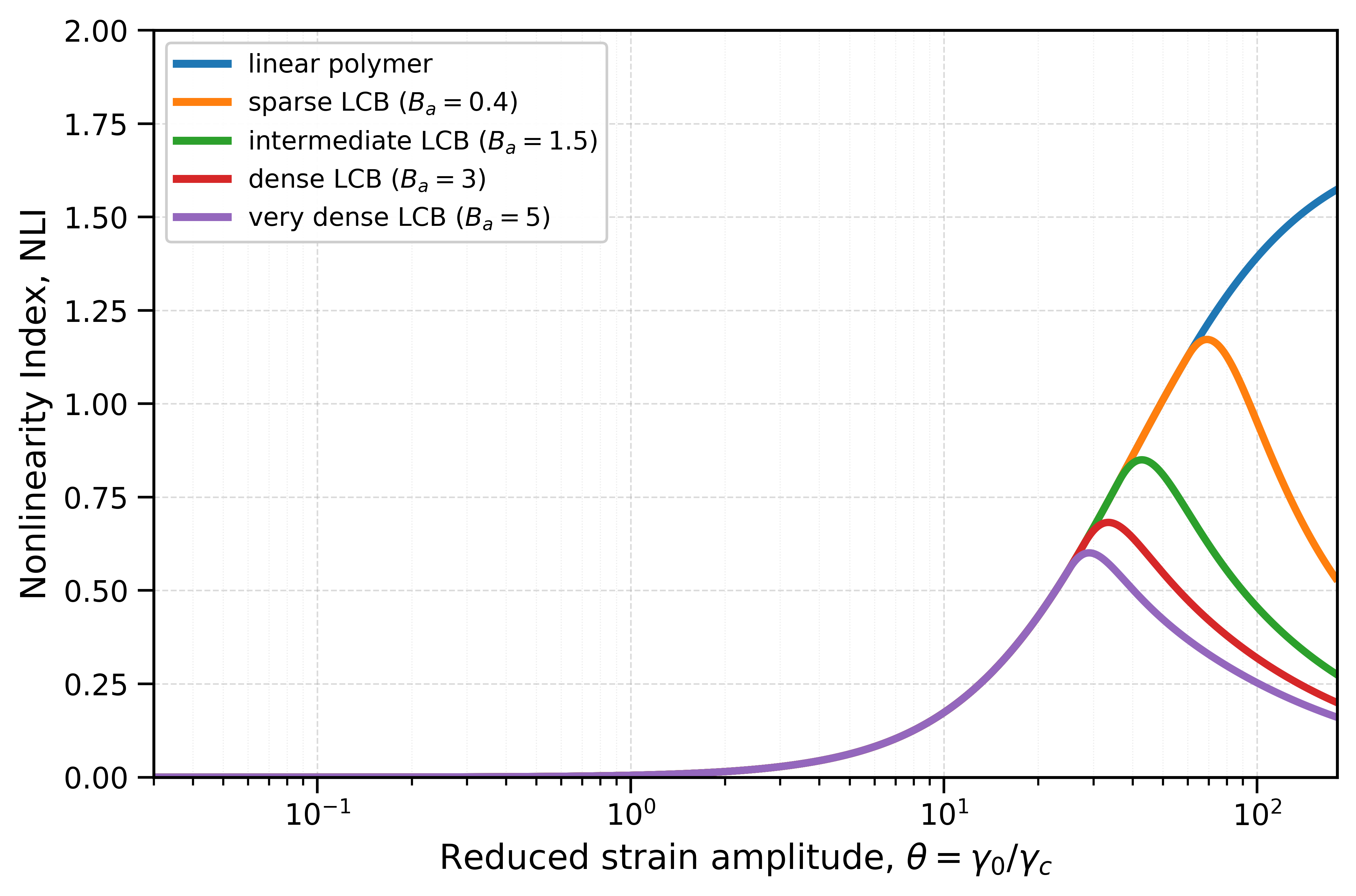}
\caption{
Theoretical prediction of the Nonlinearity Index obtained by combining the backbone nonlinear crossover response of Eq.~\eqref{eq:nli_paper1}, corresponding to Eq.~(83) of Ref.~\cite{NichettiZaccone2026Nonaffine}, with the delayed arm-retraction survival factor of Eq.~\eqref{eq:Harm_delayed}. All branched polymers initially follow essentially the same backbone nonlinear response because the long arms remain trapped within their confining tubes and do not yet contribute appreciably to stress relaxation. For sparse long-chain branching, arm retraction is activated only after the backbone has developed substantial nonlinear orientation, allowing the NLI to exceed unity before branch-mediated relaxation becomes dominant. Increasing the branching parameter $B_a=f_bZ_a/Z_{bb}$ activates branch-mediated relaxation at progressively lower reduced strain, producing broader and lower maxima characteristic of increasingly dense long-chain branching.
}
\label{fig:theory_curves}
\end{figure}

The model predictions are summarized in Fig. \ref{fig:theory_curves}. Increasing $Z_a/Z_{bb}$ lowers the peak NLI and enhances the post-peak suppression. The linear polymer limit, $Z_a=0$, gives monotonic growth toward the orientational ceiling. Sparse long-arm branching gives a peak that may transiently exceed unity before delayed arm retraction becomes dominant. Dense long-arm branching gives a peak below unity and a pronounced decrease at large reduced strain.

The magnitude of the post-peak suppression is controlled by the molecular branching parameter
\begin{equation}
B_a=\frac{f_bZ_a}{Z_{bb}},
\end{equation}
which measures the relative importance of arm entanglements compared with the backbone span between neighbouring branch points. Increasing $B_a$ enhances branch-mediated relaxation after the onset of arm retraction, producing progressively smaller NLI maxima, as illustrated schematically in Fig.~\ref{fig:theory_curves}. This representation provides a direct route from NLI measurements to an effective molecular descriptor of branching. 

Since
\begin{equation}
\frac{Z_a}{Z_{bb}}
=
\frac{M_a}{M_{bb}},
\end{equation}
the entanglement molecular weight cancels from the architecture ratio. A low NLI maximum should therefore not be interpreted simply as the consequence of a smaller $M_e$; it instead indicates longer arms, shorter backbone distances between neighbouring branch points, larger effective branch functionality, or a broader distribution of branch lengths.


\section{Physical regimes}

The complete nonlinear response can be organized into three regimes.

For $\gamma_0\leq\gamma_c$, the stress response is almost sinusoidal. Tube orientation remains proportional to strain amplitude, higher odd elastic harmonics are negligible, and the NLI is approximately zero.

For $\gamma_c<\gamma_0<\gamma_m$, backbone orientation becomes nonlinear. Branch points behave predominantly as temporary anchors and allow orientational memory to build during the oscillation cycle. CCR and tube deformation distort the waveform, and the NLI increases.

The two characteristic strains correspond to successive thresholds of the same internal molecular state variable, namely the backbone orientation transmitted through the T-branch. The onset strain $\gamma_c$ marks the deformation at which the backbone orientation first exceeds the linear-response limit, while $\gamma_m$ corresponds to the larger backbone orientation at which the entropic force transmitted through the T-branch becomes sufficient to activate efficient arm retraction. Thus, $\gamma_c$ and $\gamma_m$ are not independent characteristic strains but two milestones along the evolution of the same molecular coordinate. The maximum condition may be written schematically as
\begin{equation}
\left.\frac{dS_{xy}^{(bb)}}{dt}\right|_{\mathrm{orientation}}
\simeq
\left.\frac{dS_{xy}^{(bb)}}{dt}\right|_{\mathrm{arm\ retraction+CCR+tube\ dilation}}.
\end{equation}
For $\gamma_0>\gamma_m$, the entropic force transmitted through the T-branch becomes sufficiently large that arm retraction provides an efficient relaxation pathway during each oscillation cycle. As the long arms progressively release the stored branch-point tension, the T-branch becomes less effective at sustaining backbone orientation. The orientational memory accumulated during the first part of the cycle is therefore continuously relaxed before additional nonlinear orientation can develop. The stress waveform consequently becomes less distorted, the higher odd elastic harmonics decrease, and the NLI progressively falls despite the increasing applied strain amplitude.


\section{Linear-polymer limit}

The linear polymer is obtained by setting
\begin{equation}
Z_a=0.
\end{equation}
Then
\begin{equation}
\phi_b=1,
\qquad
S_{xy}^{(bb)}=S_{xy}^{lin},
\qquad
Z_{bb}=Z=\frac{M}{M_e}.
\end{equation}
The onset strain becomes
\begin{equation}
\gamma_c=C_cZ^{-1/2},
\end{equation}
and the peak bound reduces to
\begin{equation}
\NLI_{\max}^{lin}=3,
\end{equation}
as previously predicted by constraint-tube theory \cite{NichettiZaccone2026Nonaffine}.
In practice, a real linear polymer may not reach this theoretical ceiling within the experimentally accessible window. Nevertheless, this limit provides the reference from which the branched-polymer peak is reduced.

\section{Discussion}

The present theory provides a molecular explanation for one of the most distinctive
rheological signatures of long-chain branching under oscillatory shear: the appearance
of a broad maximum in the Nonlinearity Index followed by a progressive decay. The central physical mechanism is not a direct suppression of backbone orientation by branching itself, but the delayed activation of arm retraction. Before this activation occurs, the T-branch behaves as a transient branch-point constraint and the backbone develops nonlinear orientation essentially as in the corresponding linear polymer. Only after sufficient entropic force has accumulated at the branch point do the long arms begin to retract within their confining tubes, progressively relaxing the branch-point force balance and shortening the lifetime of backbone orientational memory.

This delayed crossover naturally explains why the experimental NLI curves initially resemble those of linear polymers before departing towards lower values at larger deformation. The central physical picture emerging from the present theory is therefore that the branch point acts as an active molecular force-transmission node rather than a passive geometrical junction. 
During the early stages of deformation, it transmits orientational stress from the backbone to the arms. At larger deformation, the same branch point becomes the gateway through which arm retraction releases the stored entropic tension. The observed NLI maximum is therefore the macroscopic manifestation of a competition between the rate at which nonlinear backbone orientation is created and the rate at which branch-mediated relaxation destroys the corresponding orientational memory.

A notable prediction emerging from the present model is that sparse and dense long-chain branching belong to two distinct nonlinear rheological regimes. Sparse long-chain branching behaves, initially, almost identically to the corresponding linear polymer because arm retraction remains inactive while the backbone undergoes essentially the same nonlinear orientational buildup as the linear chain. Only after sufficient branch-point tension has accumulated do the long arms begin to retract, allowing the NLI to exceed unity before branch-mediated relaxation becomes dominant. Consequently, the nonlinear elastic response can continue to increase beyond $\mathrm{NLI}=1$ before delayed arm retraction becomes effective. In contrast, dense long-chain branching activates branch-point relaxation at substantially lower reduced strain, suppressing the buildup of backbone orientation and leading to lower and broader NLI maxima.

More generally, the present formulation shows that molecular architecture does not primarily determine the initial buildup of the nonlinear response. Instead, it controls the efficiency of the relaxation pathway that becomes available once arm retraction is activated. This distinction provides a more transparent physical interpretation than treating branching simply as a reduction of the attainable orientation. In the present picture, branching modifies the dynamics of orientational memory rather than the mechanism of orientation itself.

The branching parameter
\[
B_a=\frac{f_bZ_a}{Z_{bb}}
\]
emerges as the principal molecular descriptor controlling the nonlinear response. Increasing $B_a$ corresponds to longer entangled arms, shorter backbone spans between neighbouring branch points, or larger effective branch functionality. All of these factors increase the efficiency of branch-mediated relaxation after the onset of arm retraction, leading to broader and progressively smaller NLI maxima. Since
\[
\frac{Z_a}{Z_{bb}}=\frac{M_a}{M_{bb}},
\]
the entanglement molecular weight cancels from the architecture ratio, making the model directly applicable across chemically different polymer systems.

The proposed framework also clarifies the relation between the present theory and recent nonaffine interpretations of nonlinear oscillatory rheology. Within the constraint-limited orientation picture developed previously, the nonlinear response of linear polymers is governed primarily by the finite orientational capacity of the entanglement network. The present work shows that the molecular fingerprint of long-chain branching introduces an additional architecture-specific nonaffine relaxation mechanism. Arm retraction progressively releases the orientational force stored at the branch point before the linear-chain orientational limit is reached. The decrease of the NLI therefore reflects the gradual loss of coherent backbone orientational memory rather than a simple saturation of nonlinear elasticity.

The analytical model deliberately isolates the minimal molecular ingredients responsible for the NLI maximum. A fully quantitative constitutive description would require tensorial tube dynamics including hierarchical branch-point motion, dynamic tube dilation, stretch relaxation, and realistic distributions of arm lengths and branch spacings. Nevertheless, the present formulation captures the essential physics while remaining sufficiently simple to provide direct molecular interpretation of nonlinear oscillatory measurements.

The present model predicts that the NLI maximum is controlled primarily by the delayed activation of arm retraction rather than by the onset of nonlinear backbone orientation itself. Consequently, measurements of the peak position and height provide direct experimental information on the competition between backbone orientation and arm relaxation, offering a route to infer molecular architecture from nonlinear oscillatory rheology.

\section{Conclusions}

We have developed a molecular tube theory that explains the characteristic peak of the Nonlinearity Index in long-chain-branched polymers as the consequence of a competition between nonlinear backbone orientation and delayed arm retraction. The central idea is that the branch point acts as a molecular force-transmission node. During the early stages of nonlinear deformation the long arms remain effectively trapped within their confining tubes and therefore behave as transient branch-point constraints ("orientation anchors"), allowing backbone orientation to develop almost identically to the corresponding linear polymer. At larger deformation the accumulated branch-point tension activates arm retraction, progressively relaxing the transmitted entropic force and shortening the lifetime of backbone orientational memory.

This physical picture naturally explains the experimentally observed broad maximum of the Nonlinearity Index. Unlike previous phenomenological interpretations of nonlinear Fourier rheology, the present theory attributes the NLI maximum to a specific microscopic relaxation mechanism whose activation is controlled by molecular architecture through delayed arm retraction. The peak does not represent a simple saturation of nonlinear elasticity. Instead, it marks the crossover between two competing molecular processes: the buildup of nonlinear backbone orientation and the delayed branch-mediated relaxation of that orientation through arm retraction. The subsequent decrease of the NLI therefore reflects the progressive loss of coherent backbone orientational memory during each oscillation cycle.

The theory predicts that the architecture dependence of the nonlinear response is governed primarily by the branching parameter
\[
B_a=\frac{f_bZ_a}{Z_{bb}},
\]
which measures the relative importance of arm entanglements compared with the orientable backbone span between neighbouring branch points. Increasing $B_a$ enhances branch-mediated relaxation after the onset of arm retraction and therefore produces broader and progressively smaller NLI maxima. This provides a direct molecular connection between nonlinear Fourier rheology and long-chain-branched molecular architecture.

Although intentionally minimal, the present theory identifies delayed arm retraction as the
dominant molecular mechanism governing the nonlinear harmonic response of long-chain-
branched polymers. Future extensions incorporating hierarchical branch-point dynamics,
molecular-weight distributions, and fully tensorial tube constitutive equations should allow
quantitative predictions for industrial polymer architectures while preserving the same
fundamental physical picture. More generally, the present work demonstrates that nonlinear
Fourier rheology can be interpreted directly in terms of identifiable molecular relaxation
mechanisms, making the Nonlinearity Index a molecular observable linking oscillatory
rheology to entanglement topology and branch architecture.

\section*{Acknowledgments}
 A.Z. gratefully acknowledges funding from the European Union through Horizon Europe ERC Grant number: 101043968 ``Multimech''.

\section*{Conflict of interest}
The authors declare that they have no conflict of interest.

\bibliography{references}

\end{document}